\documentclass[mathleft
]{an}
\usepackage{graphicx}
\usepackage{times}
\begin{document}

\Pagespan{475}{}
\Yearpublication{2009}%
\Yearsubmission{2008}%
\Month{11}%
\Volume{}%
\Issue{}%

\title{Planetary transit observations at the University Observatory Jena: XO-1b and TrES-1\thanks{Based on observations obtained with telescopes of the University Observatory Jena, which is operated by the Astrophysical Institute of the Friedrich-Schiller-University Jena.}}

\author{St. Raetz\inst{1}\fnmsep\thanks{Corresponding author:
  \email{straetz@astro.uni-jena.de}\newline}
\and M. Mugrauer\inst{1}
\and T. O. B. Schmidt\inst{1}
\and T. Roell\inst{1}
\and T. Eisenbeiss\inst{1}
\and M. M. Hohle\inst{1,4}
\and N. Tetzlaff\inst{1}
\and M. Va{\v n}ko\inst{1}
\and A. Seifahrt\inst{1,2}
\and Ch. Broeg\inst{3}
\and J. Koppenhoefer\inst{5}
\and R. Neuh{\"a}user\inst{1}
}
\titlerunning{Transit observation at the University Observatory Jena}

\institute{
Astrophysikalisches Institut und Universit{\"a}ts-Sternwarte Jena, Schillerg{\"a}{\ss}chen 2-3, 07745 Jena, Germany
\and 
Institut f{\"u}r Astrophysik, Georg-August-Universit{\"a}t, Friedrich-Hund-Platz 1, 37077 G{\"o}ttingen, Germany
\and 
Space Research and Planetary Sciences, Physikalisches Institut, University of Bern, Sidlerstra{\ss}e 5, 3012 Bern, Switzerland
\and
Max Planck Institute for Extraterrestrial Physics, Giessenbachstra{\ss}e, 85748 Garching, Germany
\and
University Observatory Munich, Scheinerstrasse 1, 81679 M\"unchen, Germany}

\received{2008 Dec 9}
\accepted{2009 Mar 30}
\publonline{2009 May 30}


\abstract{We report on observations of transit events of the transiting planets XO-1b and TrES-1 with a 25\,cm telescope of the University Observatory Jena. With the transit timings for XO-1b from all 50 available XO, SuperWASP, Transit Light Curve (TLC)-Project- and Exoplanet Transit Database (ETD)-data, including our own $I$-band photometry obtained in March 2007, we find that the orbital period is $P=(3.941501\pm 0.000001)\,$d, a slight change by $\sim$3\,s compared to the previously published period. We present new ephemeris for this transiting planet.\\ Furthermore, we present new $R$-band photometry of two transits of TrES-1. With the help of all available transit times from literature this allows us to refine the estimate of the orbital period: $P=(3.0300722\pm 0.0000002)\,$d. \\Our observations will be useful for future investigations of timing variations caused by additional perturbing planets and/or stellar spots and/or moons.} 

\maketitle

\section{Introduction}

The search for extrasolar planets is one of the most important research fields in astronomy today. Up to now the most successful method to detect exoplanet candidates is the radial velocity technique. In the last ten years a further indirect detection method has established itself as a highly successful technique in finding and confirming planets - the transit method. The first success of this technique was the confirmation of the planet candidate HD\,209458b, which was found by the radial velocity method (Mazeh et al. 2000), as a real planet by Charbonneau et al. (2000). Since that time more than 50 transiting planets have been discovered by over 20 ground-based projects and space missions.\\ The transit event is - in a first approximation - a periodic phenomenon. In a system where a known planet transits its host star, a second planet in that system will cause the time between transits to vary. This technique is itself a planet detection method that is very powerful for searching low-mass planets and is most sensitive for planets in mean-motion resonances. For example, for a hot jupiter (3-day orbit), an Earth mass planet in the 2:1 resonance will cause periodic transit timing variations that have an amplitude greater than one minute (Steffen et al. 2007). It can also be used to detect possible trojans of transiting extrasolar planets (see Ford \& Holman 2007).\\ The transit method is becoming increasingly popular, because even with small telescopes one can achieve great successes. Small ground-based observatories have already exceeded the photometric precision necessary to detect sub-Earth radius planets by the transit timing variation method (Steffen et al. 2007).\\ We have started high precision photometric observations at the University Observatory Jena in fall 2006. In this work we use the transit method to observe known transiting planets. The aim is to test procedures with our telescope and determine the obtainable photometric precision with the currently existing camera. We paid special attention to the accurate determination of transit times in order to identify precise transit timing variations that would be indicative of perturbations from additional bodies and to refine the orbital parameters of the systems. First results were presented in Va{\v n}ko et al. (2009).\\ In this paper, we describe the observations, the data reduction, and the analysis procedures. Then, we present results for XO-1b and TrES-1 that we obtained from our observations with a 25\,cm telescope at the University Observatory Jena.

\section{The Observations and Data Reduction}

All observations were carried out at the University Observatory Jena which is located close to the village Gro{\ss}schwabhausen, 10\,km west of the city of Jena.\\ Mugrauer (2009) describes the instrumentation and operation of the system in detail. In summary, there are three telescopes (90, 25 and 20\,cm) on the same mount. Because new motors for the movement of the telescope were installed recently we carry out test observations with the 25\,cm Casse- grain auxiliary telescope. We use the optical CCD-camera CTK -- \textit{Cassegrain Teleskop Kamera}. The properties of the camera are given in Table \ref{CTK}.\\ We started our observations in November 2006. Between March 2007 and July 2008, we used 46 clear nights for our transit observations. Part of the time we observed known transiting planets in Bessell $R$ and $I$ band.\\ We calibrate the images of our target fields using the standard IRAF\footnote{IRAF is distributed by the National Optical Astronomy Observatories, which are operated by the Association of Universities for Research in Astronomy, Inc., under cooperative agreement with the National Science Foundation.} procedures \textit{darkcombine}, \textit{flatcombine} and \textit{ccdproc}. We did not correct for bad pixel. Because of the large pixel scale of the camera we used (the PSF of a star consists, on average, of 3 pixels) serious errors would arise by interpolation of bad pixel with the surrounding pixels.
\begin{table}
 \centering
\caption{Camera facts of the CTK}
\label{CTK}
\begin{tabular}{lr}\hline
Fabricator: & Finger Lake \\
Type: & IMG 1024S \\
Detector: & CCD TK1024 (Tektronix) \\
Pixel: & 1024 x 1024 (24 $\mu$m) \\
Pixelscale: & (2.2065 $\pm$ 0.0008)$''$/Pixel \\
Field of View: & 37.7$'$ x 37.7$'$ \\
Filter: & Bessell $B, V, R, I$, Gunn $z$ \\
Focus: & Cassegrain \\
\hline
\end{tabular}
\end{table}

\section{Photometry and detrending}

First we perform aperture photometry. We use the IRAF task \textit{chphot} written by Christopher Broeg which is based on the standard IRAF routine \textit{phot}. With \textit{chphot} it is possible to do the photometry on every star in the field in each frame.\\ A problem in differential photometry is the search for a good comparison star. Broeg et al. (2005) developed an algorithm which uses as many stars as possible (all available field stars) and calculate an artificial comparison star. The algorithm decides which stars are the best by taking the weighted average of them. Then it computes the artificial comparison star with the best possible signal-to-noise ratio by automatically weighting down the stars according to their variability.\\ As a second step, we select the best comparison stars. We successively sort out all stars with low weights (stars that are not on every image, stars with low signal-to-noise ratio (S/N) and variable stars). As result, we get the most constant stars with the best S/N in the field.\\ In a third step we correct for systematic effects by using the Sys-Rem detrending algorithm which is proposed by Tamuz et al. (2005) and implemented by Johannes Koppenhoefer. However, the usage of Sys-Rem is not possible in every case. While observing at the University Observatory Jena, we take a large number of data points during the previously known transit time and fewer points outside. If too few out-of-transit data points before and after the transit event (constant light) are available, the transit itself is identified as sytematic effect and is removed.\\ Finally, we search for unknown variable stars in the field.

\section{XO-1b}

The first success of the \textit{XO-Project} was the discovery of the exoplanet XO-1b which transits the 11\,th magnitude, G1\,V star GSC 02041-01657 (XO-1) in the constellation Corona Borealis.\\ The \textit{XO}-cameras monitored many thousands of bright stars between September 2003 and September 2005 (detailed description in McCullough et al. 2005). XO-1 has been identified as one of dozens of stars with a transit shaped light curve. With the help of additional photometry and spectroscopy in summer 2005, the planetary nature of the transit could be confirmed (McCullough et al. 2006).

\subsection{Observation and analysis}

\begin{figure}[h]
  \centering
  \includegraphics[width=0.48\textwidth]{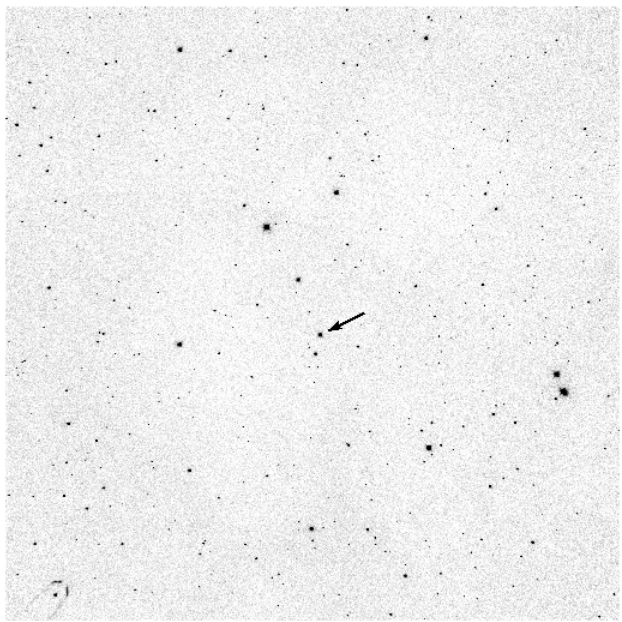}
  \caption{The 37.7$'$ x 37.7$'$ $I$-band image of the CTK. The arrow marks the star XO-1. North is up; east is to the left.}
  \label{FoV_XO1}
\end{figure}

We observed the transit of the exoplanet XO-1b in the night of 2007 March 11/12. Fig. \ref{FoV_XO1} shows the field of view. According to the ephemeris provided by McCullough et al. (2006),
\begin{equation}
\begin{array}{r@{.}lcr@{.}l}
T_{\mathrm{c}}(E)=(2453808 & 9170 & + & E\cdot 3 & 941534)\,\mathrm{d} \\
\pm0 & 0011 &  & \pm0 & 000027
\end{array}
\end{equation}
this transit corresponds to epoch 92.\\ We took 161 $I$ band exposures between 10:43 p.m. and 2:47 a.m. (UT). With an exposure time of 60\,s we achieve a mean cadence of the data points of 1.4\,min (readout time of the CCD-camera around 24\,s). Our mean photometric precision is 0.008\,mag. Due to thin clouds and fog, the photometric accurancy at the beginning of the night was a little bit worse than at the end.
\\ After calibrating all images, we perform aperture photometry on all available field stars. We found 461 stars - XO-1 and 460 comparison stars. We used an aperture of radius 8 pixels (17.65$''$) and an annulus for sky subtraction ranging in radius from 15 to 20 pixels, centered on each star.\\ To get the best possible result for the transit light curve, we try to use only the best comparison stars in the field. After a first run of the artificial-comparison-star-algorithm we reject 168 comparison stars, that could not be measured on every image. After a rerun of the algorithm we reject faint stars with low S/N and variable stars which could introduce disturbing signals to the data. With the remaining 28 objects we calculated the artificial comparison star. Finally this artificial comparison is compared to XO-1 to get the differential magnitudes for every image.\\In the resulting light curves of XO-1 and the comparison we used Sys-Rem. The algorithm works without any prior knowledge of the effects. The number of effects that should be removed from the light curves is selectable and can be set as a parameter. Already using Sys-Rem with two effects the transit itself is removed from the light curve of XO-1. Thus we use only one effect. We did not notice major changes in the resulting light curve, because the shape of this transit is not very much influenced by systematic effects. We could not correct for the thin clouds at the beginning of the night.

\subsection{Determination of the midtransit time}

\begin{figure}[h]
  \centering
  \includegraphics[width=0.48\textwidth]{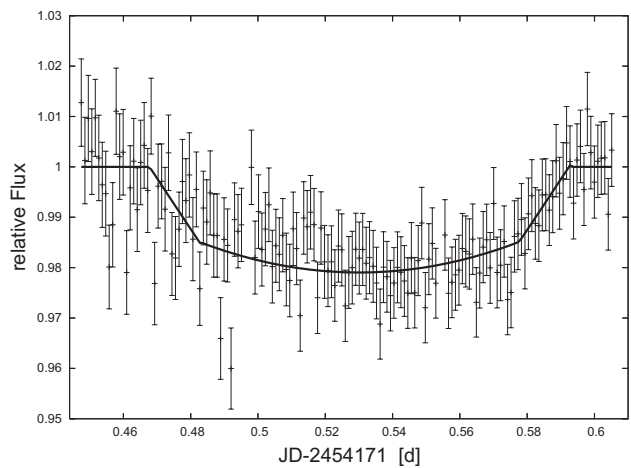}
  \caption{Relative $I$-band photometry of XO-1 from 2007 March 11/12. The solid line shows the best-fitting model.}
  \label{Lichtkurve_XO1_Flux}
\end{figure}

Because the output of the IRAF task \textit{phot} are magnitudes we converted our data to relative flux. We then calculated a weighted average of the pre-ingress and post-egress data and divided all data points by this value, in order to normalize the flux.\\ To determine the time of the transit center of XO-1 we fit a theoretical light curve to the observed light curve. We used the system parameters by Holman et al. (2006) (see Table \ref{XO1-Parameter}) to calculate this theoretical light curve of XO-1. To get the best fit we compare the theoretical light curve with the observed light curve until the $\chi^{2}$ is minimal. With the help of the $\chi^{2}$-test we could determine the time of the midtransit to $T_{\mathrm{c}}\mathrm{(HJD)}=(2454171.53258\pm0.00170)\,\mathrm{d}$. We give the 1-$\sigma$ error bars.\\ The final time series is plotted in Fig. \ref{Lichtkurve_XO1_Flux}.
\begin{table}
 \centering
\caption{System parameter for XO-1 from Holman et al. (2006)}
\label{XO1-Parameter}
\begin{tabular}{cc}\hline
Parameter & Median Value  \\
\hline
Transit Depth & 2.1\,\% \\
$t_{\mathrm{I}}$\,--\,$t_{\mathrm{IV}}$ & 2.992\,h \\
$t_{\mathrm{II}}$\,--\,$t_{\mathrm{III}}$ & 2.286\,h \\
$t_{\mathrm{I}}$\,--\,$t_{\mathrm{II}}$ & 0.353\,h \\
\hline
\end{tabular}
\end{table}

\subsection{Ephemeris}

\begin{figure}[h]
  \centering
  \includegraphics[width=0.48\textwidth]{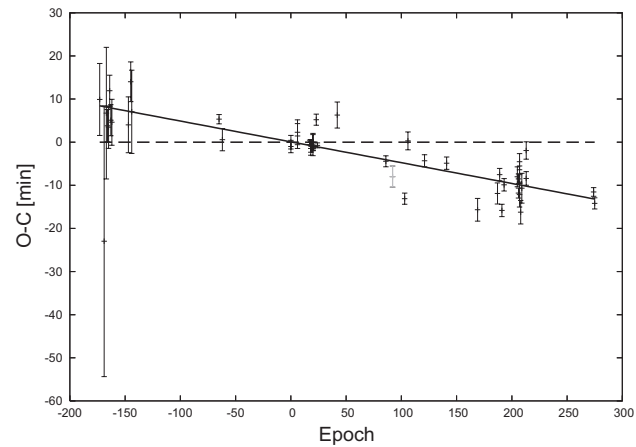}
  \caption{Transit timing residuals for XO-1. The data point from the University Observatory Jena is shown in gray. the The dashed line shows the ephemeris given by McCullough et al. (2006) The best-fitting line (solid line) is plotted, representing the updated ephemeris given in equation \ref{ephemeris_new_XO1}.}
  \label{O_C_XO1}
\end{figure}

In addition to the one transit observed at the University Observatory Jena we could find 50 transit times from 2004-2009 in the literature. These altogether 51 transits are summarized in Table \ref{Transit_times_XO1}.
\begin{table}
\caption{Summary of all known transit times of XO-1.}
\label{Transit_times_XO1}
\begin{tabular}{lcr@{\,$\pm$\,}l}
\hline
Observer & Epoch$^{a}$ & \multicolumn{2}{c}{HJD (Midtransit)} \\ \hline
XO-Survey$^{b}$ & -173 & 2453127.03850 & 0.00580 \\
& -169 & 2453142.78180 & 0.02180 \\
SuperWASP-North$^{b}$ & -167 & 2453150.68550 & 0.01060 \\
& -166 & 2453154.62500 & 0.00260 \\
& -165 & 2453158.56630 & 0.00340 \\
& -164 & 2453162.51370 & 0.00250 \\
& -163 & 2453166.45050 & 0.00250 \\
& -162 & 2453170.39170 & 0.00370 \\
& -147 & 2453229.51430 & 0.00450 \\
& -145 & 2453237.40430 & 0.00320 \\
& -144 & 2453241.34100 & 0.00670 \\
ETD$^{c}$ & -65 & 2453552.72099 & 0.00075 \\
& -62 & 2453564.54230 & 0.00180 \\
& 0 & 2453808.91630 & 0.00100 \\
XO-Survey$^{d}$ & 0 & 2453808.91700 & 0.00110 \\
ETD$^{c}$ & 6 & 2453832.56720 & 0.00200 \\
& 6 & 2453832.56780 & 0.00200 \\
TLC-Project$^{e}$ & 17 & 2453875.92305 & 0.00036 \\
& 18 & 2453879.86400 & 0.00100 \\
ETD$^{c}$ & 18 & 2453879.86360 & 0.00110 \\
TLC-Project$^{e}$ & 19 & 2453883.80565 & 0.00018 \\
& 20 & 2453887.74679 & 0.00015 \\
ETD$^{c}$ & 20 & 2453887.74700 & 0.00060 \\
& 20 & 2453887.74770 & 0.00140 \\
& 20 & 2453887.74720 & 0.00170 \\
& 23 & 2453899.57590 & 0.00090 \\
& 24 & 2453903.51330 & 0.00040 \\
& 42 & 2453974.46580 & 0.00210 \\
& 86 & 2454147.88585 & 0.00089 \\
This work & 92 & 2454171.53258 & 0.00170 \\
ETD$^{c}$ & 103 & 2454214.88590 & 0.00090 \\
& 106 & 2454226.71981 & 0.00144 \\
& 121 & 2454285.83962 & 0.00097 \\
& 141 & 2454364.66991 & 0.00100 \\
& 169 & 2454475.02536 & 0.00183 \\
& 187 & 2454545.97560 & 0.00170 \\
& 189 & 2454553.86170 & 0.00100 \\
& 191 & 2454561.73900 & 0.00100 \\
& 193 & 2454569.62620 & 0.00100 \\
& 205 & 2454616.92590 & 0.00160 \\
& 206 & 2454620.86710 & 0.00080 \\
& 206 & 2454620.86480 & 0.00080 \\
& 207 & 2454624.80610 & 0.00200 \\
& 207 & 2454624.81140 & 0.00130 \\
& 207 & 2454624.80930 & 0.00140 \\
& 208 & 2454628.74480 & 0.00190 \\
& 209 & 2454632.69020 & 0.00240 \\
& 213 & 2454648.45790 & 0.00110 \\
& 213 & 2454648.46240 & 0.00140 \\
& 274 & 2454888.88930 & 0.00070 \\
& 275 & 2454892.82900 & 0.00090 \\
\hline
\end{tabular}
$^{a}$ according to the ephemeris of McCullough et al. (2006)\\
$^{b}$ from Wilson et al. (2006)\\
$^{c}$\,from various observers collected in Exoplanet Transit Database, http://var.astro.cz/ETD \\
$^{d}$ from McCullough et al. (2006)\\
$^{e}$ from Holman et al. (2006)
\end{table}
\\ We used the ephemeris of McCullough et al. (2006) to compute ''observed minus calculated'' (O-C) residuals for all 51 transit times. Fig. \ref{O_C_XO1} shows the differences between the observed and predicted times of midtransit, as a function of epoch. The dashed line represents the ephemeris given by McCullough et al. (2006). We found a negative trend in this O-C-diagram. Thus, we refine the ephemeris using the linear function for the heliocentric Julian date of midtransit (Equation \ref{Elements}) where epoch $E$ is an integer and $T_{\mathrm{0}}$ the midtransit to epoch 0.
\begin{equation}
\label{Elements}
T_{\mathrm{c}}=T_{0}+P\cdot E
\end{equation}
For an exact determination of the orbital period we plotted the midtransit times over the epoch and did a linear $\chi^{2}$-fit. We obtained the best $\chi^{2}$ with \begin{center}
$T_{0}(\mathrm{HJD})\,=\,(2453808.91705 \pm 0.00009)\,\mathrm{d}$ \end{center} and an orbital period of \begin{center} $P\,=\,(3.941501\,\pm\,0.000001)\,\mathrm{d}$.\end{center} Within the error bars 55\,\% of the points are consistent with our calculated period and 25\,\% deviate less than 2-$\sigma$ from the new ''zero''-line (solid line in Fig. \ref{O_C_XO1}).\\ The resulting new ephemeris which represent our measurements best is \begin{equation}
\label{ephemeris_new_XO1}
T_{\mathrm{c}}\mathrm{(HJD)}\,=\,(2453808.91705\,+\,E\cdot 3.941501)\,d\end{equation}

\section{TrES-1}

TrES-1 discovered by Alonso et al. (2004) was the first success among the many ground-based, wide-field surveys for transiting planets with bright parent stars.\\ The TrES-1 parent star GSC 02652-01324 is a relatively bright ($V$\,=\,11.79\,mag) K0\,V star. It was observed with two telescopes of the Trans-Atlantic Exoplanet Survey (TrES) network (Alonso et al. 2007) in 2003. Because only one of these two telescopes could observe several transits of TrES-1 the orbital period has to be close to an integral number of days. The planetary nature of the Jupiter-sized companion was confirmed via multicolor photometry and precise radial velocity measurements.

\subsection{Observation and analysis}

We observed two transits of TrES-1 in front of it's parent star with our 25\,cm Cassegrain telescope. The first light curve from 2007 March 15 includes 88 $R$-band 60\,s exposures between 1:22 a.m. and 3:32 a.m. (UT). Unfortunately we stopped our observation to early so that the egress is missing. This transit corresponds to epoch 326 of the ephemeris given by Winn et al. (2004):
\begin{equation}
\label{Ephemeris_Winn}
\begin{array}{r@{.}lcr@{.}l}
T_{\mathrm{c}}(E)=(2453186 & 80603 & + & E\cdot 3 & 0300737)\,\mathrm{d} \\
\pm0 & 00028 &  & \pm0 & 0000013
\end{array}
\end{equation}
Our mean photometric precision of the $V$\,=\,11.79\,mag star is 0.009\,mag.\\ More than one year later on 2008 July 14 we observed a second transit event of TrES-1 at the University Observatory Jena. Because of better weather conditions the 146 $R$-band 60\,s exposures have a mean photometric precision of 0.007\,mag. \\ After data reduction we performed aperture photometry of all field stars with an aperture of 5 pixels (11.0$''$). We substract the underlying sky value, after estimating its brightness within an annulus centered on each star ranging from 15 to 20 pixels in radius.
\begin{figure}[h]
  \centering
  \includegraphics[width=0.48\textwidth]{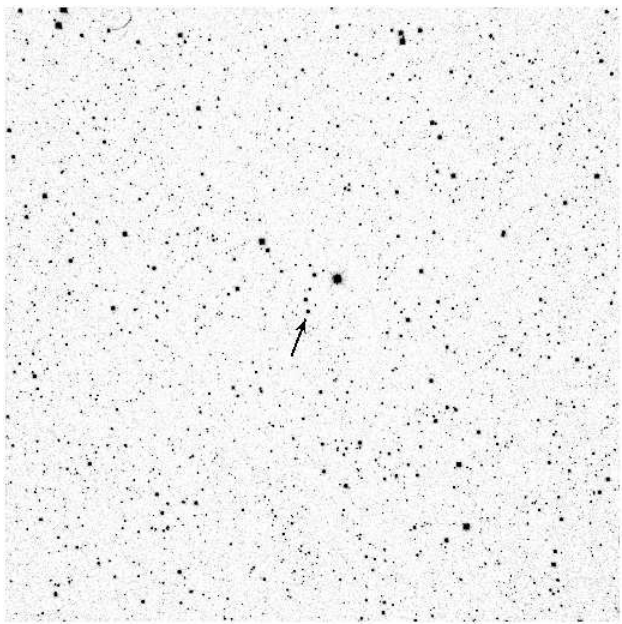}
  \caption{The 37.7$'\,\times$\,37.7$'$ CTK-field around the TrES-1 parent star observed in $R$-band. The arrow shows the star with the known transiting planet TrES-1. North is up; east is to the left.}
  \label{FoV_TrES1}
\end{figure}
\\ In the field around the TrES-1 parent star (Fig. \ref{FoV_TrES1}) we found much more objects than around XO-1. We could identify 1857 stars. For the first transit from 2007 March 15 281 stars could not be measured on every image. We reject these stars after the first run of the artificial-comparison-star-algorithm. Finally we used the 34 most constant stars with a good S/N to calculate the artificial comparison star magnitudes. For the second transit from 2008 July 14 we rejected 450 objects in the first step. With 15 constant stars we calculate the artificial comparison star.\\ We also tried to use Sys-Rem for both light curves of TrES-1. However, all 88 images from 2007 March 15 were taken during the transit, so that no constant light was available to search for systematic effects. Therefore we did our final analysis without using Sys-Rem. For the second transit it was again not possible to use Sys-Rem because of too few constant light. But we saw an obvious trend in the light curves of all stars in the field around TrES-1. We correct this effect from the data by fitting this trend and calculating the residuals.

\subsection{Determination of the midtransit time}

\begin{figure}[h]
  \centering
  \includegraphics[width=0.48\textwidth]{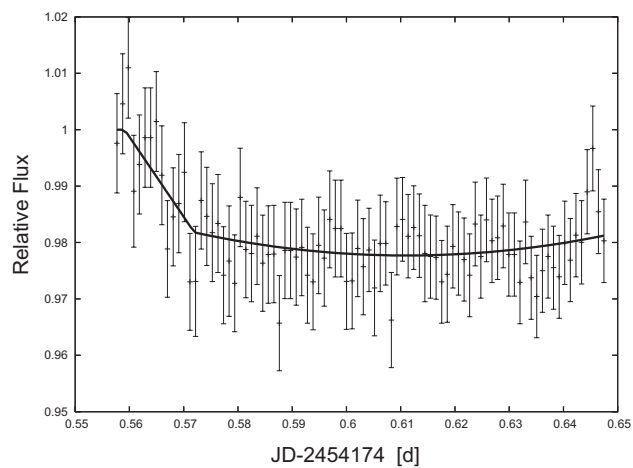}
  \caption{Relative $R$-band photometry of TrES-1 from 2007 March 15. The best-fitting model is shown as a solid line}
  \label{Lichtkurve_TrES1_Flux_2007}
\end{figure}

\begin{figure}[h]
  \centering
  \includegraphics[width=0.48\textwidth]{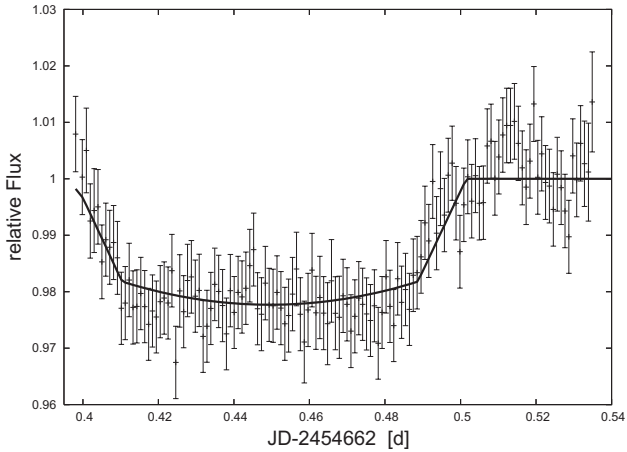}
  \caption{Relative $R$-band photometry of TrES-1 from 2008 July 14. The best-fitting model is again shown as a solid line}
  \label{Lichtkurve_TrES1_Flux_2008}
\end{figure}

We again have to convert the magnitudes to relative flux. The determination of the transit center is done the same way as in the case of XO-1. After normalization of the flux we fit a theoretical light curve using the system parameters by Winn et al. (2007) (Table \ref{TrES1-Parameter}) on our observed light curve.
\begin{table}
 \centering
\caption{System parameter for TrES-1 from Winn et al. (2007)}
\label{TrES1-Parameter}
\begin{tabular}{cc}\hline
Parameter & Median Value  \\
\hline
Transit Depth & 2.23\,\% \\
$t_{\mathrm{I}}$\,--\,$t_{\mathrm{IV}}$ & 2.497\,h \\
$t_{\mathrm{II}}$\,--\,$t_{\mathrm{III}}$ & 1.880\,h \\
$t_{\mathrm{I}}$\,--\,$t_{\mathrm{II}}$ & 0.3085\,h \\
\hline
\end{tabular}
\end{table}
\\With the help of the theoretical light curve and the $\chi^{2}$-test it was possible to determine the center of the transit even without the egress of the first transit. At a midtransit time of \\ $T_{\mathrm{c}}\mathrm{(HJD)}=(2454174.60958\pm0.00150)\,\mathrm{d}$\\ the $\chi^{2}$ is minimal. The time of the center of the second transit event could be determined to\\ $T_{\mathrm{c}}\mathrm{(HJD)}=(2454662.45165\pm0.00100)\,\mathrm{d}$. The resulting light curves are shown in Fig. \ref{Lichtkurve_TrES1_Flux_2007} and Fig. \ref{Lichtkurve_TrES1_Flux_2008}.

\subsection{Timing residuals}

\begin{figure}[h]
  \centering
  \includegraphics[width=0.48\textwidth]{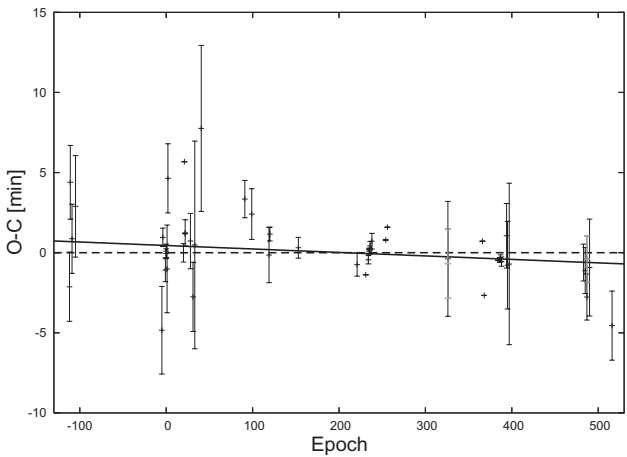}
  \caption{Transit timing residuals for TrES-1. The data points from the University Observatory Jena are shown in gray. The dashed and best-fitting (solid) line shows the ephemeris given by Winn et al. (2007) and the updated ephemeris given in equation \ref{Ephemeris_neu_TrES1}, respectively.}
  \label{O_C_TrES1}
\end{figure}

For TrES-1 we found 48 midtransit times between 2003 and 2009 in the literature. These altogether 50 transits are summarized in Table \ref{Transit_times_TrES1}. The transit of epoch 40.5 (forced e\,=\,0) according to the ephemeris given by Winn et al. (2007) is even a secondary transit observed with the \textit{Spitzer Space Telescope} by Charbonneau et al. (2005).
\begin{table}
\caption{Summary of all known transit times of TrES-1.}
\label{Transit_times_TrES1}
\begin{tabular}{lcr@{\,$\pm$\,}l}
\hline
Observer & Epoch$^{a}$ & \multicolumn{2}{c}{HJD (Midtransit)} \\ \hline
TrES network$^{b}$ & -112.0 & 2452847.43630 & 0.00150 \\
& -111.0 & 2452850.47090 & 0.00160 \\
& -109.0 & 2452856.52860 & 0.00150 \\
& -105.0 & 2452868.65030 & 0.00220 \\
& -5.0 & 2453171.65230 & 0.00190 \\
& -4.0 & 2453174.68640 & 0.00040 \\
& -1.0 & 2453183.77520 & 0.00050 \\
TrES network$^{c}$ & 0.0 & 2453186.80600 & 0.00020 \\
TrES network$^{b}$ & 0.0 & 2453186.80610 & 0.00030 \\
& 1.0 & 2453189.83540 & 0.00190 \\
& 2.0 & 2453192.86940 & 0.00150 \\
& 20.0 & 2453247.40750 & 0.00040 \\
ETD$^{d}$ & 21.0 & 2453250.44152 & \\
& 22.0 & 2453253.46848 & \\
& 22.0 & 2453253.46851 & 0.00057 \\
& 28.0 & 2453271.64860 & 0.00120 \\
& 31.0 & 2453280.73640 & 0.00150 \\
& 33.0 & 2453286.79880 & 0.00450 \\
Spitzer$^{b} $ & 40.5 & 2453309.52940 & 0.00360 \\
ETD$^{d}$ & 91.0 & 2453462.54506 & 0.00081 \\
& 99.0 & 2453486.78500 & 0.00110 \\
& 119.0 & 2453547.38470 & 0.00120 \\
& 120.0 & 2453550.41568 & 0.00030 \\
& 153.0 & 2453650.40752 & 0.00045 \\
& 221.0 & 2453856.45180 & 0.00050 \\
& 231.0 & 2453886.75210 & \\
TLC-Project$^{e}$ & 234.0 & 2453895.84297 & 0.00018 \\
& 235.0 & 2453898.87341 & 0.00014 \\
ETD$^{d}$ & 235.0 & 2453898.87350 & \\
TLC-Project$^{e}$ & 236.0 & 2453901.90372 & 0.00019 \\
Narita$^{f}$ & 238.0 & 2453907.96407 & 0.00034 \\
ETD$^{d}$ & 254.0 & 2453956.44530 & \\
& 256.0 & 2453962.50600 & \\
& 326.0 & 2454174.60979 & 0.00249 \\
This work & 326.0 & 2454174.60958 & 0.00150 \\
ETD$^{d}$ & 366.0 & 2454295.81350 & \\
& 368.0 & 2454301.87130 & \\
& 384.0 & 2454350.35400 & \\
Hrudkova$^{g}$ & 386.0 & 2454356.41417 & 0.00010 \\
& 387.0 & 2454359.44431 & 0.00015 \\
& 388.0 & 2454362.47424 & 0.00020 \\
ETD$^{d}$ & 394.0 & 2454380.65580 & 0.00140 \\
& 395.0 & 2454383.68460 & 0.00190 \\
& 397.0 & 2454389.74480 & 0.00350 \\
& 483.0 & 2454650.33120 & 0.00080 \\
& 485.0 & 2454656.39100 & 0.00100 \\
This work & 487.0 & 2454662.45165 & 0.00100 \\
ETD$^{d}$ & 487.0 & 2454662.45000 & 0.00100 \\
& 490.0 & 2454671.54150 & 0.00210 \\
& 516.0 & 2454750.32090 & 0.00150 \\
\hline
\end{tabular}
$^{a}$according to the ephemeris of Winn et al. (2007)\\
$^{b}$ from Charbonneau et al. (2005)\\
$^{c}$ from Alonso et al. (2004)\\
$^{d}$\,from various observers collected in Exoplanet Transit Database, http://var.astro.cz/ETD \\
$^{e}$ from Winn et al. (2007) \\
$^{f}$ from Narita et al. (2007) \\
$^{g}$ from Hrudkova et al (2009) \\
\end{table}
\\ With all available transit times in Table \ref{Transit_times_TrES1} we determined the transit and secondary-eclipse timing residuals for TrES-1. The calculated times (using the ephemeris of Winn et al. 2007) have been subtracted from the observed times. The resulting O-C is shown in Fig. \ref{O_C_TrES1}.\\ A slight negative trend is visible in the O-C-diagram. Using a linear $\chi^{2}$-fit we refined the ephemeris to equation \ref{Ephemeris_neu_TrES1}.
\begin{equation}
\label{Ephemeris_neu_TrES1}
\begin{array}{r@{.}lcr@{.}l}
T_{\mathrm{c}}(E)=(2453186 & 806341 & + & E\cdot 3 & 0300722)\,\mathrm{d} \\
\pm0 & 000051 &  & \pm0 & 0000002
\end{array}
\end{equation}

\section{Summary and Conclusions}

Using the University Observatory Jena with a 25\,cm Casse- grain telescope equipped with the optical CCD camera CTK we could observe the known planetary transits of XO-1b and TrES-1.\\ XO-1 was the first transit observed with our telescope at the University Observatory Jena. We determined that the orbital period is lower than previously expected. Thus our measurement of the transit time with an precision of about 1.9\,min, lead to a refined estimate of the ephemeris.\\ The second star with a known transiting planet observed in our observatory was TrES-1. Despite the missing egress of the first transit event we could determine the center of the transit. With this observation of two transits of TrES-1 we could refine the previously published ephemeris. \\ The mean photometric precision of the two observed stars ($V$\,=\,11\,-\,12\,mag) with known transiting extrasolar planets is 0.008\,mag and the precision in the determination of the transit times is $\approx$\,0.0013\,d. This allows us to register transit time variations of around 150\,s. The transit observations with our telescope at the  University Observatory Jena provide anchors for future searches for transit time variations. \\ During our transit observations we found several variable star candidates in the fields around XO-1 and the TrES-1 parent star. Since only one and two nights of observations were available, respectively, the type of variability and the orbital elements are still under investigation as successfully done for a W UMa-variable in the field around TrES-2 (Raetz et al. 2009).

\acknowledgements
The authors would like to thank the GSH Observer Team for the nightly observations. SR and MV acknowledge support from the EU in the FP6 MC ToK project MTKD-CT-2006-042514. RN acknowledges general support from the German National Science Foundation (Deutsche Forschungsgemeinschaft, DFG) in grants NE 515/13-1, 13-2, and 23-1. AK acknowledges support from DFG in grant KR 2164/8-1. TOBS acknowledges support from Evangelisches Studienwerk e.V. Villigst. TR would like to thank the DFG for financial support (grant NE 515/23-1). TE and MMH acknowledge support from the DFG in SFB-TR 7 Gravitational Wave Astronomy. Moreover we thank the technical staff of the University Observatory Jena, especially Tobias B{\"o}hm.\\This work has used data obtained by various observers collected in Exoplanet Transit Database, http://var.astro.cz/ETD. Some of the mid-transit times in this publication are from listings in the Amateur Exoplanet Archive.

\end{document}